\documentclass[trackchanges,twocolumn]{aastex701}

\usepackage{multirow}

\newcommand{\AH}[1]{#1}

\begin{document}

\title{The fraction of periodic SN Ib/c light curves}

\author[orcid=0009-0000-6748-4319,sname='Horowicz']{Asaf Horowicz}
\affiliation{{Department of Particle Physics and Astrophysics, Weizmann Institute of Science, Rehovot 76100, Israel}}
\email[show]{asaf.horowicz@weizmann.ac.il}  
\author[orcid=0000-0002-6786-8774,sname='Ofek']{Eran O. Ofek}
\affiliation{{Department of Particle Physics and Astrophysics, Weizmann Institute of Science, Rehovot 76100, Israel}}
\email[hide]{eran.ofek@weizmann.ac.il}
\author[orcid=0000-0002-3653-5598,sname='Gal-Yam']{Avishay Gal-Yam}
\affiliation{{Department of Particle Physics and Astrophysics, Weizmann Institute of Science, Rehovot 76100, Israel}}
\email[hide]{avishay.gal-yam@weizmann.ac.il}

\begin{abstract}
Periodic luminosity modulations have been recently identified for the stripped-envelope supernovae SN~2022jli and SN~2022esa, motivating a systematic search for a similar behavior in a larger sample. Such modulations may indicate the explosions arise in binary progenitor systems. We perform the first systematic search for periodic modulation in a sample of 34 Type Ib/c supernovae with high quality photometry from the Zwicky Transient Facility.
We develop and apply a statistically rigorous pipeline for detecting periodic modulation in light curves. The pipeline successfully recovers the previously reported periodic undulations of SN~2022jli and SN~2022esa, and identifies SN~2020sgf as an additional promising periodic candidate. Injection-recovery simulations are used to quantify the survey sensitivity as a function of period and modulation amplitude.
Comparing the observed detections with recent population synthesis models shows that, under the adopted assumptions, models predicting intrinsic periodic fractions of order $\sim20\%$ are consistent with the observations. \AH{Given that this inference rests on only three dual-band detections out of 34 candidates, we regard this fraction as illustrative rather than a precise measurement.} Our results suggest that periodically modulated SN 2022jli-like events may represent a \AH{non-negligible} sub-population of stripped-envelope supernovae rather than being exceptionally rare events, while demonstrating a methodology suitable for future wide-field transient surveys.

\end{abstract}


\section{Introduction}
\label{sec:intro}

Recent observations of supernovae (SNe), such as those of SN 2022jli, have shown that some events exhibit periodic modulations in their luminosity during the decline phase \citep{Moore+23,Chen+24,Maeda+25,2015ap}. 
At least in some cases, such periodicity provides indirect evidence for the presence of a binary companion to the SN progenitor, as suggested by \citet{Chen+24}.
The fact that three cases (SN 2022jli, SN 2022esa, SN 2015ap) of such behavior were observed in Type Ib/Ic SNe supports this hypothesis, as many of these SNe are thought to originate from progenitor stars that are part of binary systems, where the companion stripped the outer hydrogen and helium layers of the progenitor star before its core collapsed \citep[e.g.][]{Prentice+19, Taddia+18, Drout+23}. Recent population synthesis work investigated the interaction of a newly-born compact object and its shocked companion star, namely Compact-Companion Interactions (CCI), and concluded that up to about 30\% of stripped-envelope SNe (SESNe) might show that behavior \citep{Ercolino+26}. In this work we aim to systematically search for similar signals across a sample of \AH{34} SESNe and assess the prevalence of this phenomena.
In Section \ref{sec:methods} we describe the data set and the methodology used to search for periodic modulation. In Section \ref{sec:results} we present the results of the search. In Section \ref{sec:discussion} we discuss the survey sensitivity and compare our findings with theoretical population models to constrain the intrinsic fraction of periodic SESNe. Finally, we summarize our conclusions in Section \ref{sec:summary}.

\section{Methods} \label{sec:methods}

\subsection{Data Selection and Pre-processing}
\label{subsec:data_selection}
Using the Transient Name Server (TNS\footnote{\url{https://www.wis-tns.org}}), we selected SNe up to September 16, 2025, filtering by $z\le0.05$ and types Ib, Ic, Ib/c, Ic-BL, Ibn. We also include Ia-CSM objects due to the recent 2022esa event, that was initially misclassified as Ia-CSM and later reclassified as Ic-CSM \citep{Maeda+25}. We fetched Zwicky Transient Facility \citep[ZTF;][]{ZTF+19} $g-$ and $r-$ bands light curves (LCs) for candidates from the Lasair broker \citep{Smith+19} and retrieved light curves of 438 candidates. The photometry is based on forced photometry or subtraction images \citep{Masci+18}, where the subtraction is performed using the ZOGY algorithm \citep{Zackay+16}.

To ensure sufficient statistical leverage for our modeling procedure, we required a minimum of 20 detections per band; consequently, a periodicity search was conducted only in a single band for some of the candidates. This threshold is motivated by the structure of our fitting framework (Section~\ref{sec:model_fitting}). In the most flexible configuration, we fit a sixth-degree polynomial together with two additional sinusoidal parameters, corresponding to a total of nine free coefficients. With 20 data points, this leaves at least 11 effective degrees of freedom in the fit, ensuring that the model is meaningfully constrained.

Applying the selection criteria described above leaves a total of \AH{117} light curves: 68 belonging to 34 candidates with dual-band photometry and \AH{49} belonging to candidates with photometry in only a single band. We therefore adopt the 34 dual-band candidates as our final sample.

\subsection{Common approaches}

Typically, when searching for periodicities in a light curve, the global decline trend is fitted and subtracted. The residuals left are then searched for periodicity. A common way to search for periodicity is using a Lomb-Scargle periodogram \citep{Lomb+76, Scargle+82}. If a preferred frequency is found, the next step is to estimate its False Alarm Probability (FAP) - how likely it is to obtain a power as high, or higher, than the power of that frequency, if the entire data is just noise. While widely used and powerful for identifying periodicity in individual objects, we highlight two caveats of this approach, particularly in the context of a survey-wide search for periodicity across many light curves:

\begin{itemize}{
\item  The trend and periodic components are fitted sequentially rather than simultaneously. In practice, the data is first modeled with a polynomial baseline, and the residuals are then searched for periodicities. Consider a signal of the form $y(t)=at^2+bt+c+A\sin(2\pi ft)+B\cos(2\pi ft)$. If the sinusoidal terms are even partially correlated with the polynomial baseline -- for example when only a small number of cycles are sampled -- then fitting and subtracting the polynomial first may absorb part of the periodic signal. As a result, the residuals no longer contain the full sinusoidal component, and the subsequent period search can underestimate its amplitude or significance.

\item The interpretation of FAP may be tricky. Even for a single light curve, the precise meaning of the Lomb–Scargle FAP can be subtle, as it depends on the frequency range searched and the effective number of independent trials. Extending this to a large survey of hundreds of light curves further complicates the interpretation, since one must properly account for the total number of frequencies and targets examined in order to estimate a meaningful global FAP for entire sample.
}\end{itemize}

To address these issues while maintaining a clear statistical interpretation, we adopt a straightforward modeling approach in which the polynomial baseline and sinusoidal terms are fitted simultaneously at each trial frequency. By adjusting all parameters together, we avoid constraining the periodic component through prior detrending and naturally account for correlations between the trend and the oscillatory terms. In addition, this framework allows us to define the statistical significance of each detection in a transparent and well-controlled manner, directly from the improvement in $\chi^2$ associated with the inclusion of the sinusoidal terms. \AH{We directly compare our joint-fitting approach against this sequential detrend-then-search approach in App.~\ref{sec:gls_comparison}.}

\subsection{Model and Fitting Procedure}
\label{sec:model_fitting}

For each light curve we extract the decline detections, starting from \AH{the time of} peak magnitude \footnote{There are three cases with double peaks -- SN 2019vsi, SN 2022xxf, \AH{and SN 2022jli} -- in those cases the decline was assumed to begin after the second peak.} and up to either the first gap of 50 days, or until an epoch of 1000 days after the peak. We \AH{assume the decline can be described by a smooth function of time and} model the magnitude as

\begin{equation}
m(t) = A \sin(2\pi f t)
       + B \cos(2\pi f t)
       + \sum_{k=0}^{d} c_k \tilde{t}^k ,
\end{equation}

where:
\begin{itemize}
    \item $d$ is the polynomial degree,
    \item \AH{$c_k$ are the polynomial coefficients,}
    \item $\tilde{t}$ is the time coordinate shifted to zero mean and scaled to the interval $[-1,1]$ for numerical stability,
    \item $f$ is a trial frequency,
    \item $A$ and $B$ parameterize the sinusoidal component.
\end{itemize}

The polynomial degree is chosen to be 3 for the entire sample, to avoid overfitting signals by higher-degree polynomials. Results obtained with other baselines are shown in the Appendix \ref{sec:appendix_band_table}, Table \ref{tab:all_baselines}.

\begin{figure}
    \centering
    \includegraphics[width=0.48\textwidth]{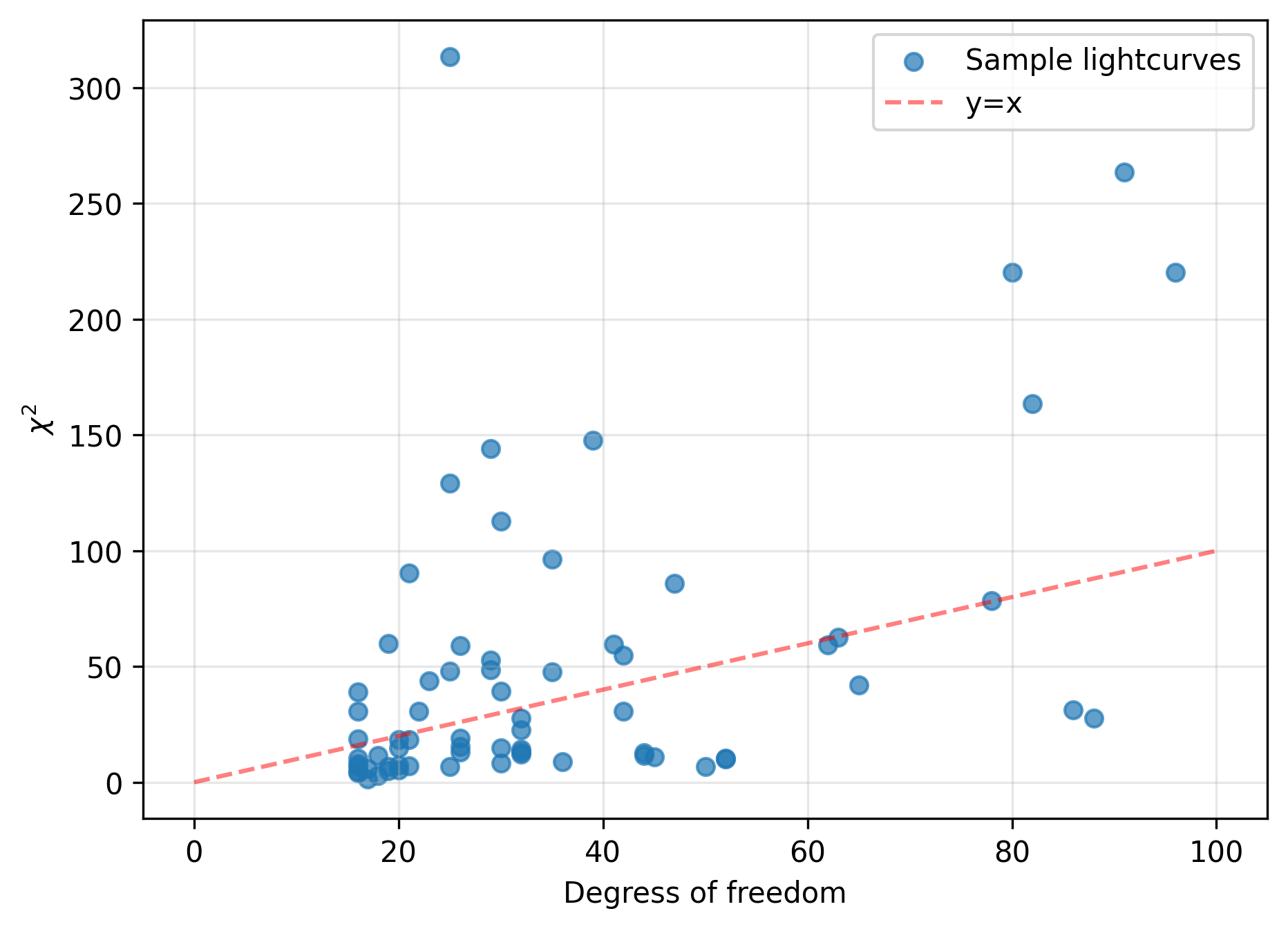}
    \caption{$\chi^2$ of the polynomial baseline fit as a function of the number of degrees of freedom for the 68 analyzed light curves. The dashed line indicates the expectation $\chi^2\sim\nu$. The uncertainties include an additional $0.015$ mag systematic term added in quadrature to the reported photometric uncertainties. Many light curves nevertheless lie above the expected relation, indicating that the adopted uncertainties underestimate the observed scatter. For these cases, the uncertainties are rescaled by a factor $\sqrt{\chi^2_\nu}$ before performing the periodicity search, see Sect. \ref{sec:model_fitting}.}
    \label{fig:chi2_dof}
\end{figure}

For each trial frequency we solve for the coefficients ${c_k, A, B}$ using weighted linear least squares. Specifically, we minimize
$$\chi^2 = \sum_i \frac{\big(y_i - m(t_i)\big)^2}{\sigma_i^2},$$
where $\sigma_i$ denotes the uncertainty assigned to each measurement, computed as the reported photometric uncertainty added in quadrature with a $0.015$ mag term accounting for possible systematic uncertainties in the ZTF photometry. Figure~\ref{fig:chi2_dof} shows that many light curves still have $\chi^2>1$, indicating that the quoted uncertainties underestimate the observed scatter. For those cases, we compute the reduced chi-square, $\chi^2_\nu=\chi^2/\nu$, where $\nu=N-k$ is the number of degrees of freedom, with $N$ the number of data points and $k$ the number of fitted polynomial coefficients. Then we rescale all uncertainties by a factor $\sqrt{\chi^2_\nu}$ and repeat the fit. This procedure prevents inflation of $\chi^2$ - it enforces $\chi^2\approx\nu$ for the baseline fit and ensures that, under the assumption of Gaussian and independent residuals, the resulting statistic is consistent with a $\chi^2$ distribution with $\nu$ degrees of freedom.

The frequency grid is defined individually for each LC according to its time baseline $T = t_{\max} - t_{\min}$. 
We restrict the search to periods that are both physically meaningful and sufficiently sampled within the data.

In principle $P = T/2$ is the longest meaningful searchable period, corresponding to two full observed cycles. In practice this is not sufficiently restrictive when fitting a third-degree polynomial baseline. A cubic polynomial can produce one maximum and one minimum, and therefore can mimic approximately one full oscillation across the data span. If only two cycles are present, the polynomial may effectively absorb one of them, leaving only a single cycle to constrain the sinusoidal component. In this regime, long-period sinusoids can resemble smooth polynomial curvature, potentially yielding artificially improved fits at large periods. To reduce this degeneracy, we conservatively restrict the search to $P \leq 2.5\ T$:
\begin{equation}
f_{\min} = \frac{5}{2T}.
\end{equation}

The minimum searchable period is set to $P_{\min}=3~\mathrm{days}$,  but theoretically could be shorter. This choice is motivated by the typical ZTF cadence of 1--2 nights and by the population synthesis models of \citet{Ercolino+26}, which predict few CCI events with periods below $\sim3$ days (their Fig.~7).  The corresponding maximum frequency is

\begin{equation}
f_{\max} = \frac{1}{3} \ \left[\mathrm{days}^{-1}\right].
\end{equation}

The frequency grid spacing is chosen as
\begin{equation}
\Delta f = \frac{1}{5T},
\end{equation}
which oversamples the frequency resolution $1/T$ by a factor of five. 
This choice ensures that the peaks in $\Delta\chi^2(f)$ are resolved while keeping the number of effectively independent frequencies of order $(f_{\max}-f_{\min})T$.

\subsection{Statistical Significance}
\label{subsec:methods_significance}
To evaluate the presence of a periodic signal, we compare two nested models:

\begin{itemize}
    \item Null model (baseline only):
    \begin{equation}
    m_{\mathrm{poly}}(t) = \sum_{k=0}^{d} c_k \tilde{t}^k
    \end{equation}

    \item Alternative model (baseline + sinusoid):
    \begin{equation}
    m_{\mathrm{full}}(t) = m_{\mathrm{poly}}(t) + 
    A \sin(2\pi f t) + B \cos(2\pi f t)
    \end{equation}
\end{itemize}

For each frequency we compute

\begin{equation}
\Delta \chi^2 = 
\chi^2_{\mathrm{poly}} - \chi^2_{\mathrm{full}}.
\end{equation}

Since the alternative model introduces two additional parameters ($A$ and $B$) per frequency, we expect

\begin{equation}
\Delta \chi^2_f \sim \chi^2_{2}
\end{equation}

under the null hypothesis of no periodic modulation. 
The single-frequency $p$-value is therefore
\begin{equation}
\label{eq:pval}
p_f = 1 - F_{\chi^2(2)}(\Delta \chi^2_f),
\end{equation}
where $F_{\chi^2(2)}$ is the cumulative distribution function of the $\chi^2$ distribution with two degrees of freedom.

For each light curve we record the frequency that maximizes $\Delta\chi^2$ and its corresponding period $P = 1/f$. The semi-amplitude is given by $\sqrt{A^2 + B^2}$.

A \AH{single-band ``detection''} is defined as a LC with $p_f < 10^{-4}$. This value is chosen such that the global FAP is less than (see  \AH{App.~\ref{app:global_fap_estimate}).} A \AH{``dual-band detection''} is defined as a SN for which both bands exhibit statistically significant periodicities within a tolerance of $P^2/T$ (adopting the larger $P^2/T$ value between the two bands). This tolerance corresponds to up to one cycle shift throughout the entire baseline. In practice, we first identify the strongest $\Delta\chi^2$ peak in one band and then search for a significant local maximum within the allowed tolerance in the other band. This procedure reduces the risk of missing real candidates when the global maximum in one band corresponds to a harmonic or aliases rather than the fundamental period.

\AH{This procedure rests on two main assumptions: that the adopted polynomial (plus, for the alternative model, sinusoidal) form is a reasonable description of the smooth component of the light curve, and that the noise is Gaussian and uncorrelated between epochs, so that $\Delta\chi^2_f$ follows a $\chi^2_2$ distribution under the null hypothesis (Eq.~\ref{eq:pval}). We test the Gaussianity assumption on our four individually-discussed candidates with a Shapiro--Wilk test (App.~\ref{app:shapiro_wilk}), and further validate our significance estimates per band and per candidate with a bootstrap test described in Sect.~\ref{sec:global_fap}.}

\subsection{Period uncertainty estimation}

The uncertainty on each recovered period is estimated from the local shape of the $\Delta\chi^2(f)$ curve around its maximum, $\Delta\chi^2_{\rm peak}$. We identify the interval over which the $\Delta\chi^2(f)$ curve remains within the $68\%$ confidence region, using the threshold
\begin{equation}
    \Delta\chi^2_{\rm crit}=\chi^{-1}_{2}(0.68),
\end{equation}
The lower and upper frequency bounds, $f_{\rm lo}$ and $f_{\rm hi}$, are defined by the two threshold-crossing points satisfying
\begin{equation}
    \Delta\chi^2(f)=\Delta\chi^2_{\rm peak}-\Delta\chi^2_{\rm crit},
\end{equation}
which are obtained by linear interpolation between adjacent frequency grid points. The corresponding confidence interval in period is obtained through
\begin{equation}
    P_{\rm lo}=\frac{1}{f_{\rm hi}}, \qquad
    P_{\rm hi}=\frac{1}{f_{\rm lo}},
\end{equation}
and the reported period uncertainty is derived from these bounds. 

\subsection{Minimal Detectable Amplitude}
\label{sub:mda}
In addition to the formal $p$-value criterion derived from $\Delta\chi^2$, we apply a complementary amplitude-based check to assess the robustness of candidate periodicities. We estimate the intrinsic scatter of each light curve by fitting the baseline model (without the sinusoidal component) and computing the standard deviation of the residuals,

\begin{equation}
\sigma_{\mathrm{res}} = \mathrm{std}(m - m_{\mathrm{baseline}}).
\end{equation}

an order of magnitude estimate of the smallest coherent sinusoidal modulation that could be reliably distinguished from this noise floor scales as
\begin{equation}
\mathcal{A}_{\mathrm{MDA}} 
\sim 
\frac{\sigma_{\mathrm{res}}}{\sqrt{N}}.
\end{equation}
We refer to this quantity as the \emph{minimal detectable amplitude (MDA)}. It reflects the practical sensitivity set by the data quality. For every analyzed light curve we calculate $\mathcal{A}_{\mathrm{MDA}}$. A \AH{detection} must also have a detected amplitude greater than $\mathcal{A}_{\mathrm{MDA}}$.  This allows us to distinguish between periodic signals whose amplitudes are physically meaningful relative to the intrinsic scatter of the light curve, and those who could have arisen from noise. In practice, all candidates satisfying the adopted significance threshold also exceeded the MDA criterion. The latter therefore acts as an additional conservative consistency check rather than the primary detection criterion.

\subsection{\AH{False-Alarm Probability}}
\label{sec:global_fap}

\AH{For each light curve and band, we assess the significance of the recovered periodicity with a bootstrap test \citep{Efron+94}: we resample the baseline-fit residuals 2000 times, refit the full model at the fixed best-fit frequency $f_0$, and take the FAP as the fraction of realizations whose $\Delta\chi^2$ meets or exceeds the observed one. Unlike the single-frequency $p$-value of Eq.~\ref{eq:pval} or the survey-wide estimate in App.~\ref{app:global_fap_estimate}, this test requires no assumption about the frequency grid or cadence. We report this FAP for each individually discussed candidate.}

\section{Results} \label{sec:results}

After applying the quality cuts described in Sect. \ref{subsec:data_selection}, we applied the periodicity search described in Section~\ref{sec:methods} to a total of \AH{117} light curves corresponding to \AH{83} unique SNe, 34 of them with both g- and r- band LCs.

Using a single-frequency threshold of $p < 10^{-4}$ and requiring the fitted sinusoidal amplitude to exceed the MDA, we identify \AH{two} SNe exhibiting significant periodicity in both the $g-$ and $r-$ bands (\AH{dual-band detections}\AH{: SN~2022jli and SN~2020sgf}). \AH{A third, SN~2022esa, has a $g$-band $p$-value of $1.01\times10^{-4}$, at our threshold. Minor analysis choices move it to either side (e.g., without the $0.015$ mag systematic floor, $p=9.7\times10^{-5}$). Given its significant $r$-band detection at a consistent period, its dual-band detection with the poly4 and poly5 baselines (App.~\ref{sec:appendix_band_table}), its bootstrap FAP~$<5\times10^{-4}$ in both bands (Sect.~\ref{sec:known}), and the independent report of its periodicity by \citet{Maeda+25}, we include it in our statistics, and refer to the three as dual-band detections hereafter.} For each candidate, the detected period, amplitude, and single-frequency $p$-value are listed in Table \ref{tab:double_hits_poly3}. The detailed results per band are provided in Table \ref{tab:band_level}. We start by discussing two previously known objects in Sect. \ref{sec:known}, and move on to new findings in Sect. \ref{sec:new}.

\begin{table*}
\centering
\begin{tabular}{lcccccc}
\hline
SN & $P_g$ & $A_g$ & $p_g$ & $P_r$ & $A_r$ & $p_r$ \\
& [Days] & [mag] & & [Days] & [mag] & \\
\hline
2020sgf & $30.2^{+0.3}_{-1.2}$ & 0.20 & $6\times 10^{-10}$ & $35.9^{+\AH{1.9}}_{-1.5}$ & \AH{0.11} & $\AH{3}\times 10^{-9}$ \\
2022esa & $34.2^{+2.2}_{-3.5}$ & 0.14 & $\AH{1}\times 10^{-4}$ & $34.2^{+2.7}_{-1.8}$ & 0.11 & $5\times 10^{-5}$ \\
2022jli & $12.6^{+0.1}_{-0.3}$ & 0.07 & $\AH{3}\times 10^{-12}$ & $12.4^{+0.1}_{-0.3}$ & 0.04 & $\AH{5}\times 10^{-6}$ \\
\hline
\end{tabular}
\caption{Dual-band periodic detections identified using a third-degree polynomial baseline. Periods, semi-amplitudes, and single-frequency $p$-values are listed for both bands.}
\label{tab:double_hits_poly3}
\end{table*}

\begin{table*}
\centering
\begin{tabular}{lccccc}
\hline
SN & Band & MDA & $\Delta\chi^2$ & $\chi^2_{\text{poly}}$ & $\chi^2_{\text{full}}$ \\
\hline
\multirow{2}{*}{2022jli} & g & 0.01 & \AH{53.31} & 91.00 & \AH{37.69} \\
& r & 0.01 & \AH{24.57} & 82.00 & \AH{57.43} \\
\hline
\multirow{2}{*}{2020sgf} & g & 0.02 & \AH{42.34} & 78.00 & \AH{35.66} \\
& r & 0.01 & \AH{39.24} & 96.00 & \AH{56.76} \\
\hline
\multirow{2}{*}{2022esa} & g & 0.03 & \AH{18.40} & 29.00 & \AH{10.60} \\
& r & \AH{0.01} & \AH{19.73} & 30.00 & \AH{10.27} \\
\hline
\end{tabular}
\caption{Single band periodicity diagnostics for the dual-band detections. For each photometric band, we report the minimal detectable amplitude, $\Delta\chi^2$ and the $\chi^2$ values for both the polynomial-only and full (polynomial+sinusoid) models. By construction, $\chi^2 \approx \nu$ (see Sect. \ref{sec:model_fitting}).}
\label{tab:band_level}
\end{table*}

\subsection{Previously Known Objects}
\label{sec:known}

We examine SN~2022jli, which has been reported to exhibit a periodic modulation of 12.4 days \citep[e.g.,][]{Chen+24}, as well as SN 2022esa with a period of $\sim 32$ days \citep{Maeda+25}. The proximity to the lunar synodic period is discussed in Sect. \ref{subsec:sensitivity}. \AH{Both candidates were included in our 34-object dual-band sample on the same footing as every other candidate and their recovery serves as a validation of the pipeline.}

\subsubsection{SN 2022jli}
\label{subsec:2022jli}

\begin{figure*}
    \centering
    \includegraphics[width=\linewidth]{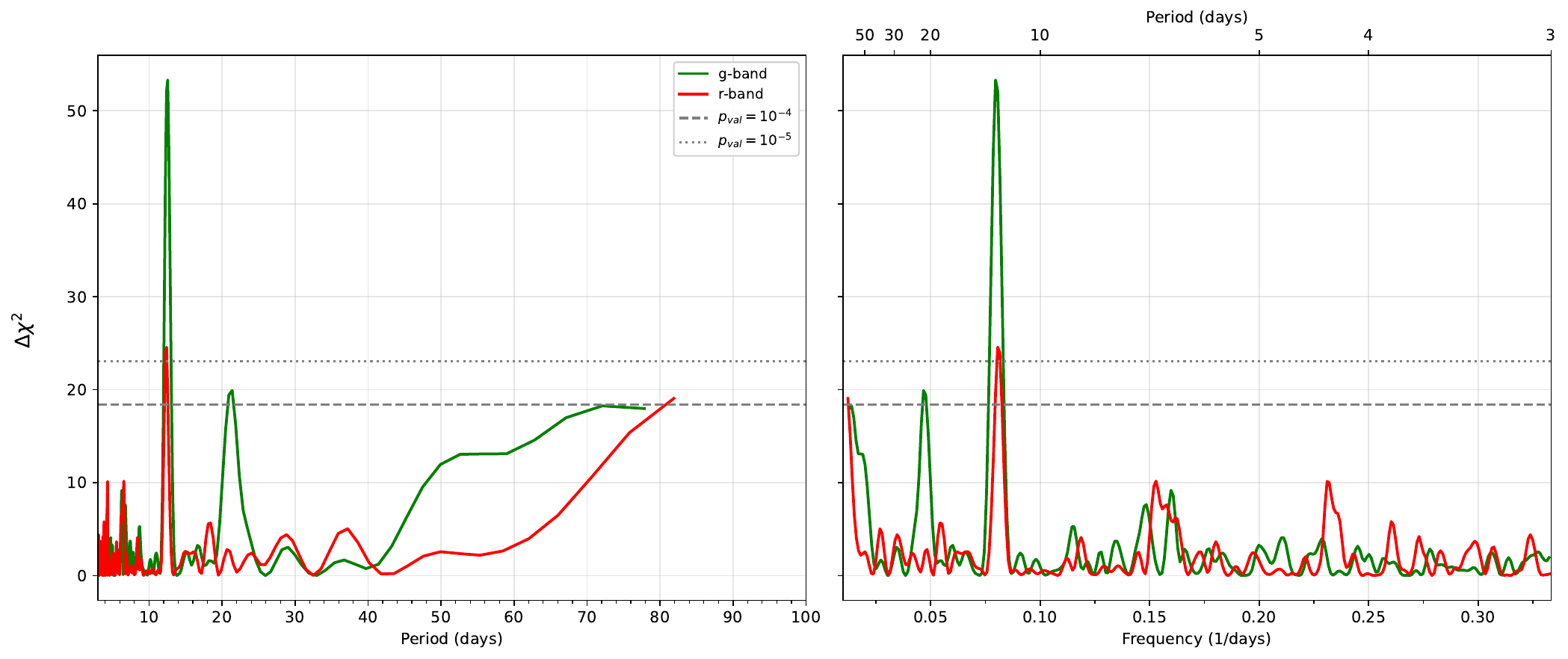}
    \caption{$\Delta\chi^2$ as a function of trial period (left) or frequency (right) for SN 2022jli. The green and red curves correspond to the two photometric bands (ZTF $g-$band, $r-$band respectively). Dashed lines are showing the $\Delta \chi^2$ values corresponding to p-values of $10^{-4},\ 10^{-5}$. Shaded green and red regions show the $68\%$ confidence intervals of the $g$ and $r$ bands respectively. A sharp and consistent peak is recovered at $P \approx 12.5$ days in both bands, in agreement with the literature. Broad structures at longer periods arise because slow sinusoids can partially compensate residual baseline mismatches, producing a gradual improvement in $\Delta\chi^2$ that very unlikely corresponds to a periodic signal.}
    \label{fig:2022jli_periodogram}
\end{figure*}

SN 2022jli was discovered by Libert Monard on 2022 May 5 (JD=2459704.67). Multiple surveys have since confirmed its detection and classified it as a SN of Type Ic. \AH{This SN was extensively studied} following the works by \cite{Moore+23} and \cite{Chen+24}\AH{, who reported periods of $12.5\pm0.2$ days and $12.4$ days, respectively, across multiple bands}. \AH{\citet{Chen+24} additionally reported $H\alpha$ emission-line shifts recurring with the same period, interpreted as evidence for periodic accretion onto a compact object.} Our pipeline detected the periodicity of SN 2022jli with high confidence ($p_f < 10^{-5}$ for both bands), \AH{recovering periods of $12.6^{+0.1}_{-0.3}$ and $12.4^{+0.1}_{-0.3}$ days for the $g$ and $r$ bands respectively (Table~\ref{tab:double_hits_poly3}), consistent with both \citet{Moore+23} and \citet{Chen+24}.} \AH{A bootstrap FAP test (as described in Sect.~\ref{sec:global_fap}) gives FAP~$<5\times10^{-4}$ in both bands.} Figure~\ref{fig:2022jli_periodogram} shows $\Delta\chi^2(f)$ for both bands.
A clear peak is observed at $P \approx 12$ days, consistent with previous analyses.

\AH{\citet{Chen+24} find the periodic signal of SN~2022jli to be asymmetric -- a fast rise over $\sim$0.25 phase followed by a slower decline -- rather than a symmetric sinusoid. We tested whether our sinusoidal model underestimates the amplitude of such a signal by injecting a sinusoid and a sawtooth-like signal into SN~2022jli's poly3 baseline (same cadence and errors, both bands), over a grid of periods (10--15 days) and semi-amplitudes (0.03--0.20 mag) bracketing SN~2022jli's own values, with 100 random-phase realizations per grid point. A trial is counted as a successful period recovery if the recovered period falls within the same $P^2/T$ tolerance used to define a detection in Sect.~\ref{subsec:methods_significance}. The period is recovered successfully in $>96\%$ of trials for both waveforms. The sinusoid's amplitude is recovered without bias (within a few percent) across essentially the entire tested grid, while the sawtooth's amplitude is underestimated by $\sim20\%$. As we discuss in Sect. \ref{subsec:ercolino_models}, this fact does not affect our inferred results.}

\subsubsection{SN 2022esa}

\begin{figure*}
    \centering
    \includegraphics[width=\linewidth]{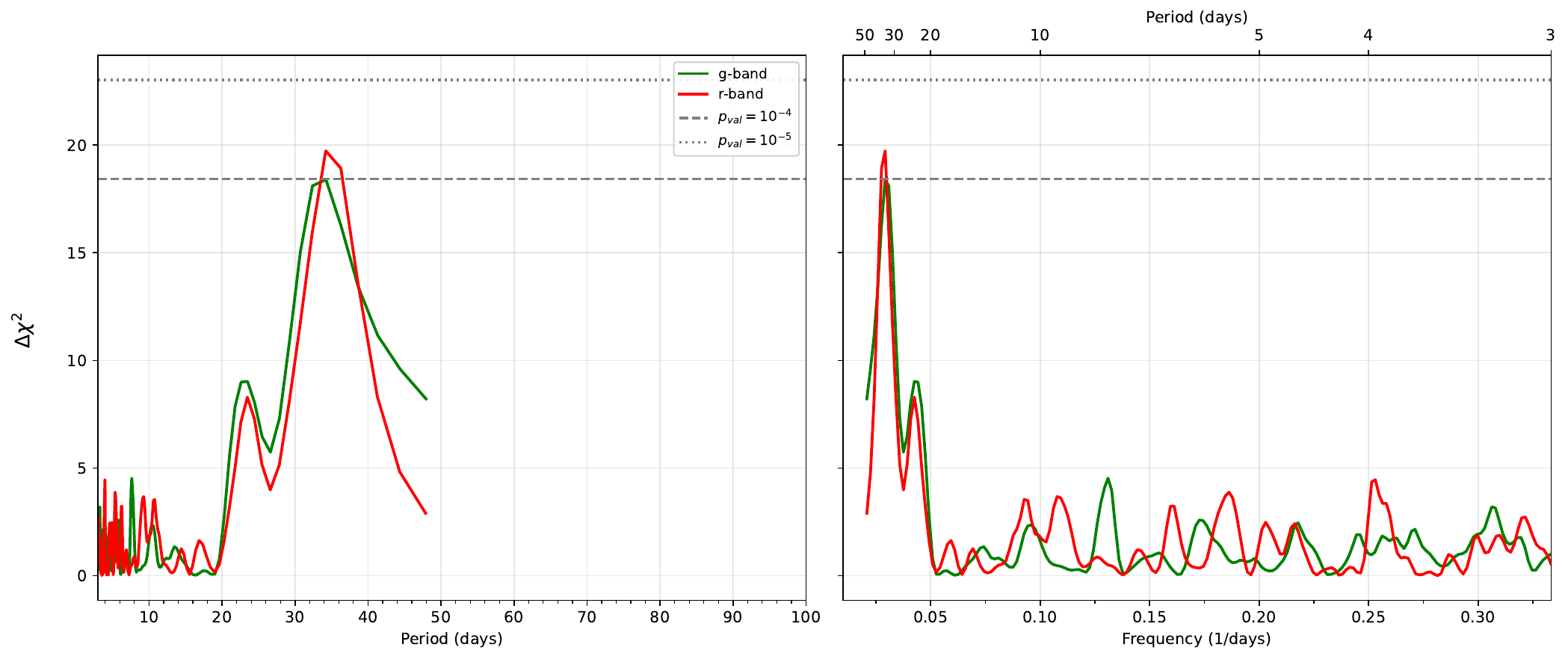}
    \caption{Same as Fig. \ref{fig:2022jli_periodogram} but for SN 2022esa. Peaks are recovered at $P=34.2$ days for both $g-$ and $r-$ bands.}
    \label{fig:2022esa_periodogram}
\end{figure*}

SN 2022esa was discovered on 2022 March 12 (JD=2459650.61) by the Asteroid Terrestrial-impact Last Alert System \citep[ATLAS;][]{Atlas+18, astronote_22esa}. It was first misclassified as a SN Ia-CSM, but later reclassified as a Ic-CSM by \cite{Maeda+25}\AH{, who showed periodic undulations of $31.0 \pm 2.2$ days ($g$ band) and $32.0 \pm 2.1$ days ($r$ band) across ZTF $g$- and $r$-bands, as well as the ATLAS $o$-band}. Our pipeline detected SN 2022esa as a \AH{dual-band detection} with a period of $34.2^{+2.2}_{-3.5}$ days for the $g$ band and $34.2^{+2.7}_{-1.8}$ days for the $r$ band.
Figure \ref{fig:2022esa_periodogram} shows the periodograms of SN 2022esa, showing a clear peak at $\approx$ 34 days. We note that there are inconsistencies of frequencies between our findings and those of \cite{Maeda+25}, and that their method is inherently different than ours - they use a 5th degree polynomial baseline, clean the residuals using $3\sigma$ clipping and perform a Lomb-Scargle analysis. They detect periods that differ by $\sim 5\%$ from ours, and also claim those periods vary throughout the decline. Importantly, both of our approaches recover similar periods, and SN 2022esa is taken into account for our final estimation of the rate of periodic behavior. \AH{A bootstrap FAP test (as described in Sect.~\ref{sec:global_fap}) gives FAP~$<5\times10^{-4}$ in both bands.}
In order to assess our results consistency, we refer to Table \ref{tab:all_baselines} to see the obtained periods for SN 2022esa with baselines of other polynomial degrees. In Sect.\ref{subsec:sensitivity} we discuss lunar synodic period effects that could be related to comparable periods.

\subsection{\AH{Newly Identified Candidate}}
\label{sec:new}

We now discuss \AH{a} newly identified candidate\AH{, SN~2020sgf,} that our pipeline found. We refer the reader to Appendix \ref{app:2019tsf} for another interesting candidate, \AH{SN~2019tsf,} that was not considered a \AH{dual-band detection} but may be interesting for further investigation.

\subsubsection{SN 2020sgf}
\label{sec:2020sgf}

\begin{figure*}
    \centering
    \includegraphics[width=\linewidth]{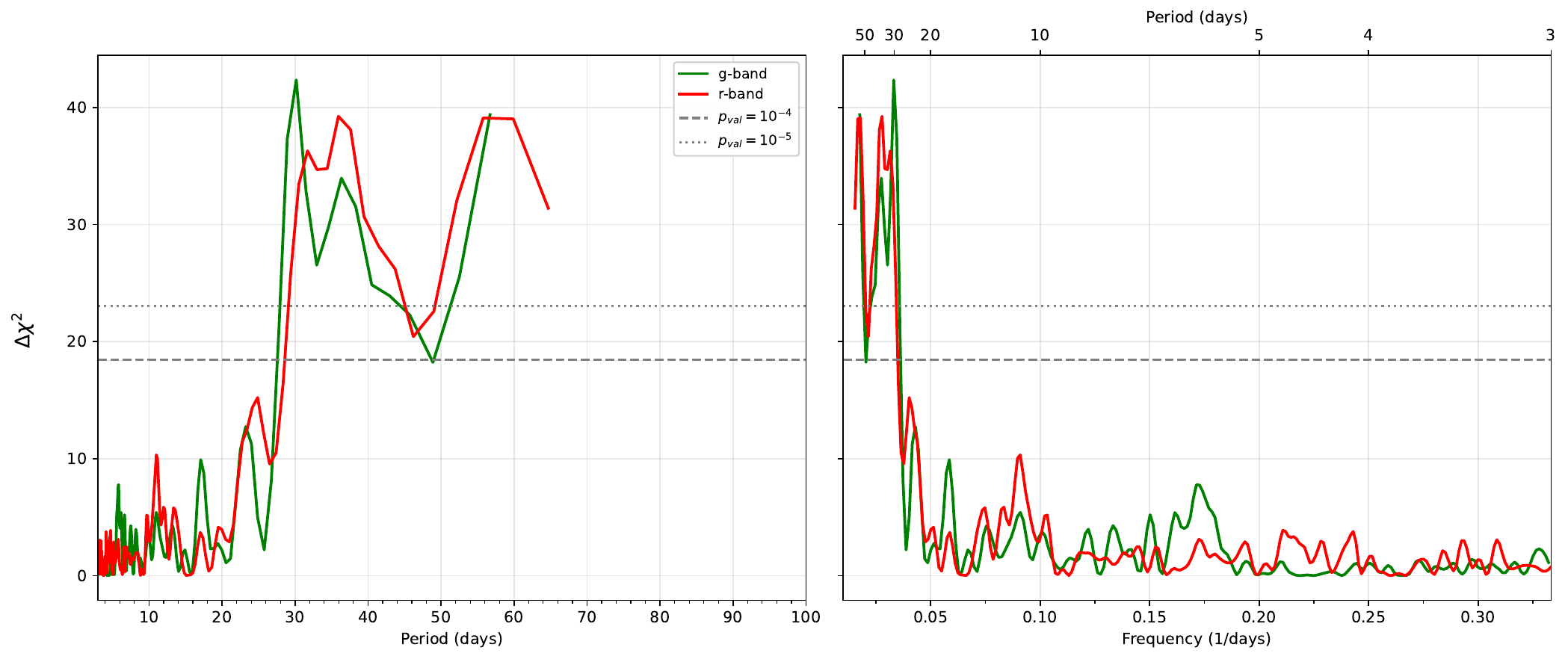}
    \caption{Same as Fig. \ref{fig:2022jli_periodogram} but for SN 2020sgf. A peak is recovered at $P \approx 30$ days for $g-$band and at $P \approx 36$ for $r-$ band. The global peak is obtained for $g-$ band, and the $r-$band peak is found within a tolerance of $P^2/T\approx8$ days, see Sect.~\ref{subsec:methods_significance}}
    \label{fig:2020sgf_periodogram}
\end{figure*}

\begin{figure*}
    \centering
    \includegraphics[width=\linewidth]{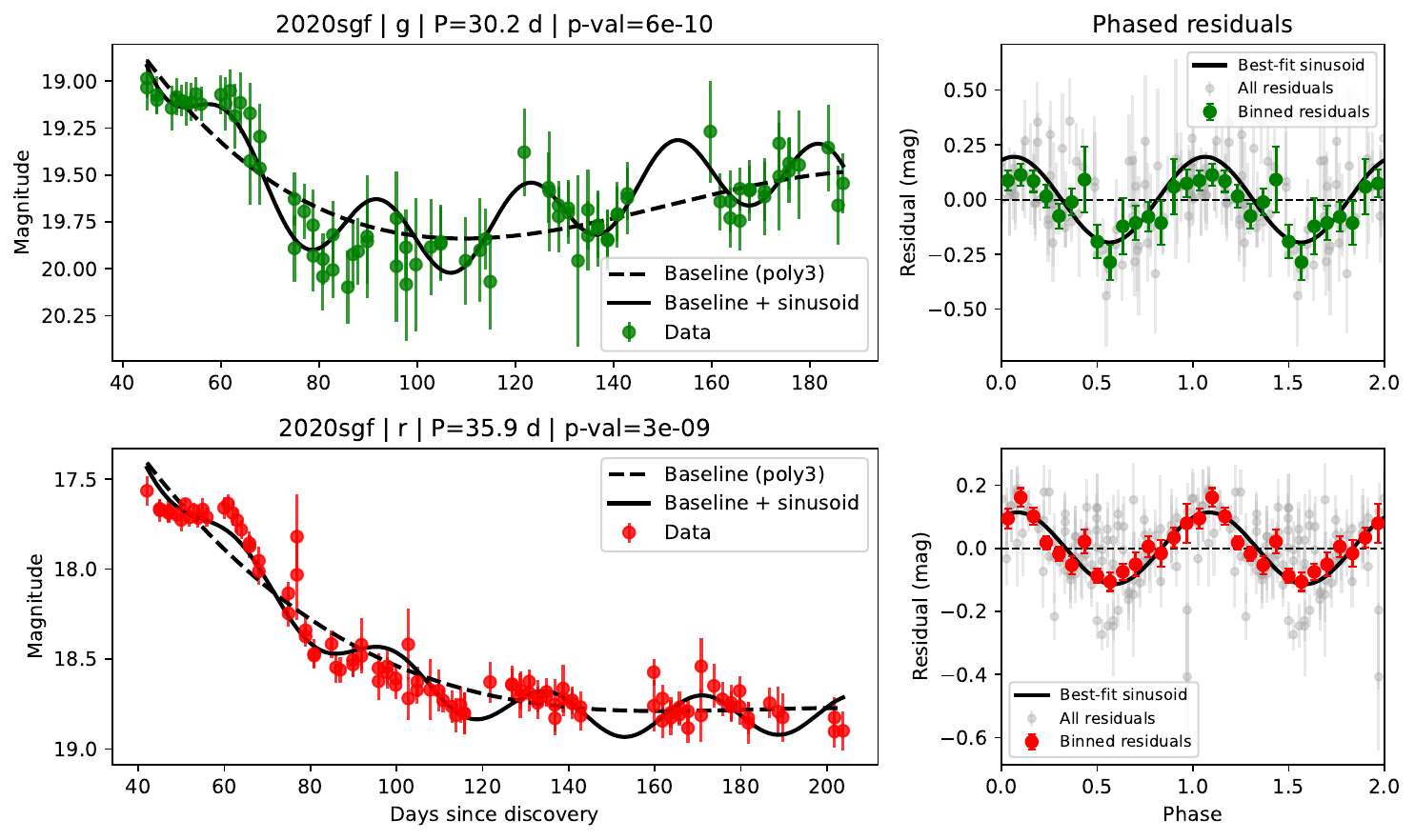}
    \caption{The decline phase of SN~2020sgf in $g$-band (upper panel) and $r$-band (lower panel). On the left panels the dashed curves show the 3rd degree polynomial baseline and the solid curves are the best-fit for the baseline + sinusoidal models. On the right - folded phase residuals obtained by subtracting the best-fit polynomial model. Phased residuals are binned in 30 equal-width phase bins. A binned point value is calculated using inverse-variance weighting, with the error on each binned point given by the standard error of the inverse-variance weighted mean.}
    \label{fig:2020sgf_fit}
\end{figure*}

A prominent result of our pipeline is the newly identified periodic candidate SN~2020sgf. It was discovered by \AH{K. De with} ZTF on 2020 August 28 \AH{(JD=2459089.95)} \citep{2020sgf_discovery} and classified as Type Ic supernova \citep{2020sgf_classification}. Our pipeline recovered a period of $\approx 30$ days in both the $g-$ and $r-$ bands. Figure \ref{fig:2020sgf_periodogram} shows the periodogram of SN~2020sgf. A clear and narrow peak at $P \approx 30$ days is present in the $g$ band and at $P\approx 36$ days in the $r$ band. With a baseline of $\approx\AH{142}$ days for the $g$ band and $\approx\AH{162}$ days for the $r$ band, those peaks correspond to period tolerances of $P^2/T\approx 6.4$ days and $P^2/T \approx 8.0$ days for $g-$, $r-$ bands respectively, therefore a tolerance of $8.0$ days is enforced and this candidate was considered a \AH{dual-band detection}. Another pair of harmonic peaks is present at $\approx 60$ days in both bands\AH{; App.~\ref{sec:gls_comparison} shows that a sequential detrend+Lomb--Scargle analysis in fact prefers this $\approx56$--$60$-day peak over the fundamental, so this candidate's period should be interpreted with some caution (see also Sects.~\ref{subsec:2020sgf_moon} and \ref{subsec:sensitivity})}. Figure \ref{fig:2020sgf_fit} shows the 3rd degree polynomial fit of the light curve of SN 2020sgf, with and without the sinusoidal period, as well as the folded phased residuals. \AH{A bootstrap FAP test (as described in Sect.~\ref{sec:global_fap}) gives FAP~$<5\times10^{-4}$ in both bands.} Further research is needed to determine the nature of SN 2020sgf. In the context of this work, we cannot rule out a periodic phenomenon and therefore include this candidate in our statistics.

\subsubsection{\AH{Possible connection to the lunar cycle}}
\label{subsec:2020sgf_moon}

\begin{figure*}
    \centering
    \includegraphics[width=\linewidth]{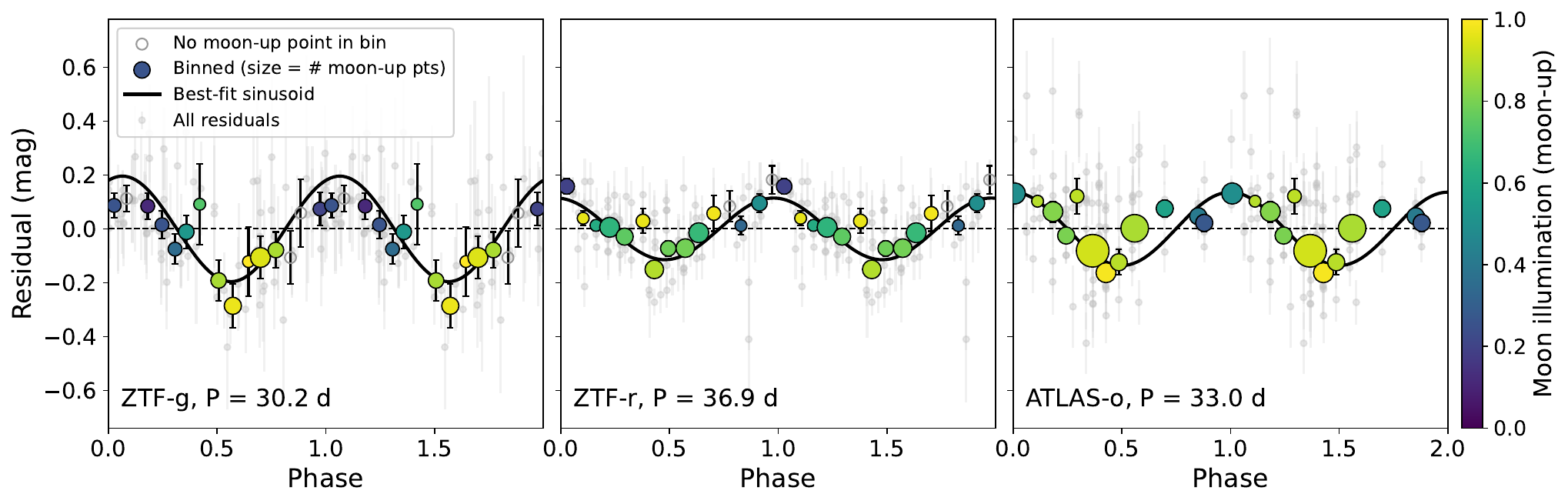}
    \caption{\AH{SN~2020sgf residuals folded on the candidate's own best-fit period (black curve: best-fit sinusoid), binned in phase and colored by the mean Moon illuminated fraction among the moon-up epochs in each bin (point size scales with the number of such epochs; hollow points have no moon-up epoch in that bin). Left to right: ZTF $g$-band, ZTF $r$-band, and an independent ATLAS $o$-band light curve of the same object, analyzed with the same pipeline. The ATLAS $o$-band period ($P\approx33$ d) is consistent with the ZTF bands and shows the same phase-illumination pattern.}}
    \label{fig:2020sgf_moon_phase}
\end{figure*}

\AH{SN~2020sgf's recovered period ($\approx30$--$36$ days) is close enough to the 29.5-day lunar synodic month that we investigated a possible connection directly, beyond the survey-wide argument in Sect.~\ref{subsec:sensitivity}. We computed the Moon's illuminated fraction and altitude above the Palomar horizon (site of the Samuel Oschin Telescope) at each ZTF epoch using \texttt{astropy}/\texttt{astroplan} \citep{astroplan}, and folded the poly3-baseline residuals on SN~2020sgf's own best-fit period, binning in phase and coloring each bin by the mean Moon illumination among only the epochs where the Moon was above the horizon. Figure~\ref{fig:2020sgf_moon_phase} shows the result for both ZTF bands together with an independent check: the ATLAS $o$-band forced-photometry light curve of the same object, fit with the same pipeline.}

\AH{In both ZTF bands and in the ATLAS $o$-band, the residuals are anti-correlated with Moon illumination: a brighter Moon coincides with the light curve appearing brighter (lower residual magnitude) than the smooth polynomial baseline, and a darker Moon with it appearing fainter. This correlation is statistically significant in each band individually, by a permutation test that shuffles the residuals against the Moon illumination 2000 times while preserving the true observing window: $g$-band, $2.5\sigma$; $r$-band, $>3.5\sigma$ (0 of 2000 shuffles as extreme as the real data); ATLAS-$o$, $2.2\sigma$. Reassuringly, an independent period search on the ATLAS $o$-band light curve recovers a consistent period of $P\approx33$ days and shows the same anti-correlation with Moon illumination (Fig.~\ref{fig:2020sgf_moon_phase}, right panel), which argues against an instrument-specific systematic (e.g., ZTF-specific image-subtraction artifacts) as the sole explanation.}

\AH{We emphasize that this is not conclusive evidence that SN~2020sgf's undulation is Moon-related. We cannot determine from the present data alone whether this indicates a real periodicity, a systematic, or some combination of both. Given this ambiguity, we treat SN~2020sgf's detection with caution but do not exclude it from our statistics, consistent with our treatment elsewhere in this work. Further investigation of this object, including a more detailed study of its relation to the lunar cycle, is left to future work.}

\subsection{Injection-Recovery Simulations}
\label{sec:injection_recovery}

To quantify the sensitivity of our search, we performed injection-recovery simulations using the observed ZTF LCs themselves. This approach preserves the actual cadence, photometric uncertainties, seasonal gaps, and time baselines of each object. For each candidate and ZTF band, we injected a synthetic sinusoidal signal directly into the measured magnitudes,
\begin{equation}
m_{\rm inj}(t)=m_{\rm obs}(t)+A\sin\left(2\pi t/P+\phi\right),
\end{equation}
where $A$ is the injected semi-amplitude, $P$ is the injected period, and $\phi$ is a random phase drawn independently for each realization.

All dual-band candidates were included in the simulations, including objects with detected periodicity, in order to characterize the sensitivity of the full analyzed sample. Since injections into detected objects are added on top of any existing variability, the recovery probabilities for those few objects may be slightly affected near their intrinsic periods. However, excluding them would cause the simulated sample to differ from the survey on which the periodicity search was performed.

The injected LCs were analyzed using the same pipeline as the real search, adopting a third-degree polynomial baseline model and the period-search procedure described in Section~\ref{sec:methods}. An injection was considered recovered only if it satisfied the full dual-band detection criterion: both bands passed the adopted single-frequency significance threshold, the recovered amplitudes exceeded the corresponding band-level MDA, and the recovered periods were mutually consistent with the injected period according to the same period-tolerance criterion used in the real search.

Recovery simulations were performed over a grid of periods and semi-amplitudes. The period bins spanned $P=5$--$65$ days in steps of $10$ days, while the semi-amplitude bins covered $A=0.00-0.20$ mag in steps of $0.02$ mag. We also tested logarithmically spaced semi-amplitude bins. Since the recovery efficiency is already very low below $\sim0.02$ mag, the increased resolution in this regime provides little additional information, and we therefore adopt uniformly spaced bins. Within each period-amplitude bin, injected periods and amplitudes were drawn uniformly, and 1000 realizations were generated for each candidate.

We define the survey sensitivity, $P_{ij}$, as the fraction of injected signals recovered in period-amplitude bin $(i,j)$,
\begin{equation}
\label{eq:pij}
P_{ij}=
\frac{N_{\rm rec}(P_i,A_j)}
     {N_{\rm max}},
\end{equation}
where $N_{\rm rec}$ is the total number of recovered injections in the bin and $N_{\rm max}=34\,000$ is the maximal number of injection realizations over the full dual-band sample (34 candidates with 1000 realizations each). Candidates whose LC baselines are too short to probe a given period bin contribute zero recovered injections. Because $P_{ij}$ is normalized by the full sample of 34 SNe, it incorporates both the recovery efficiency of the search pipeline and the fact that only a subset of objects has sufficient temporal coverage to detect signals in a given period bin. Thus, $P_{ij}$ should be interpreted as the probability that a periodic event with period $P_i$ and semi-amplitude $A_j$ would be recovered by our search pipeline if it occurred in a randomly selected SN from our sample.

\begin{figure}
\centering
\includegraphics[width=\columnwidth]{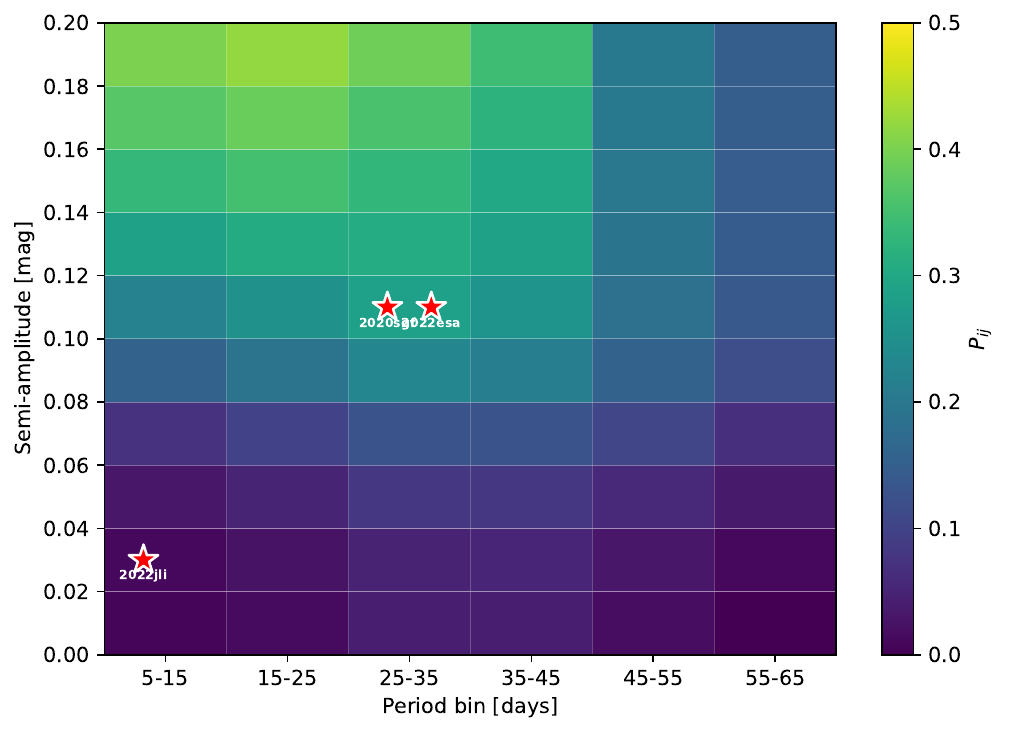}
\caption{Recovery probability map, $P_{ij}$ (Eq.~\ref{eq:pij}), obtained from the injection-recovery simulations described in Section~\ref{sec:injection_recovery}. Colors indicate the probability that a periodic signal with a given period and semi-amplitude would be recovered by our search if injected into a randomly selected SN from our sample. The three detected candidates are overplotted in the corresponding period--amplitude bins for visual reference only; their locations do not influence the values of $P_{ij}$. Sensitivity increases with increasing amplitude and decreases toward longer periods because fewer SNe have sufficiently long baselines to detect such signals.}
\label{fig:Pij_recovery_map}
\end{figure}

The resulting recovery map is shown in Figure~\ref{fig:Pij_recovery_map}. As expected, the recovery probability increases with increasing amplitude and decreases toward longer periods, reflecting the reduced number of candidates with baselines sufficiently long to probe long-period variability. The overplotted detected candidates are shown using the smaller of their measured semi-amplitudes in the two bands, reflecting the requirement that periodicity be detected independently in both bands.

\section{Discussion}
\label{sec:discussion}

\subsection{Survey reliability and sensitivity}
\label{subsec:sensitivity}

Our search identifies three \AH{dual-band periodic events} among the 34 SESNe with usable light curves in both ZTF bands: SN~2020sgf, SN~2022jli, and SN~2022esa\AH{, the latter at our significance threshold (Sect.~\ref{sec:results})}.

\begin{figure}
\centering
\includegraphics[width=\columnwidth]{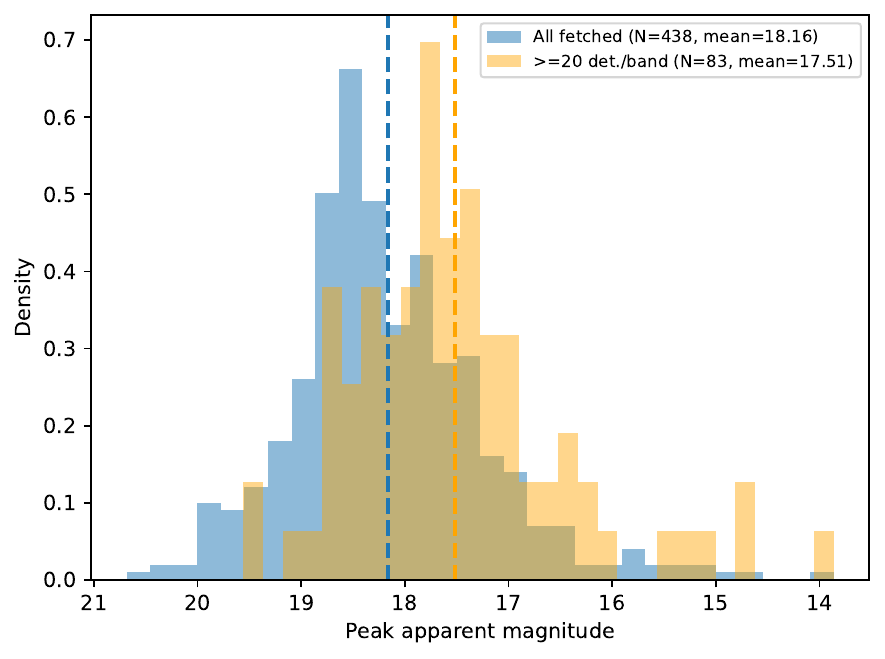}
\caption{\AH{Peak apparent magnitude distribution for all 438 TNS-selected Ib/c-like candidates with a usable ZTF light curve, already restricted to $z\le0.05$ (blue), compared with the 83 candidates (117 light curves) satisfying our $\geq20$-detections-per-band requirement (orange). Dashed vertical lines mark the sample means. The selected sample is systematically brighter, demonstrating the selection bias discussed in Sect.~\ref{subsec:sensitivity}.}}
\label{fig:sample_selection_bias}
\end{figure}

Before interpreting the detected periodic candidates, it is important to consider the selection effects of the survey. \AH{Figure~\ref{fig:sample_selection_bias} compares the peak-magnitude distribution of all 438 TNS-selected candidates with usable ZTF light curves -- already restricted to $z\le0.05$, per Sect.~\ref{subsec:data_selection} -- against the 83 objects (117 LCs) that satisfy our $\geq20$-detections-per-band requirement. The selected sample is on average brighter at peak (mean $17.5$ vs.\ $18.2$ mag) than the full candidate pool, confirming quantitatively that our sample is biased toward intrinsically brighter events. To test whether this bias drives our inferred periodic fraction, we repeated this comparison restricted to a redshift cut ($z\le0.028$) that reduces the candidate pool to roughly half its original size (221 candidates with a usable light curve, 51 satisfying our detection threshold, 22 with dual-band photometry). The 3 periodic candidates remain, out of 22 dual-band candidates in this less-biased sub-sample, still a non-negligible fraction. We emphasize that our findings, presented in the following text, cannot be easily generalized to the broader SESNe population as exact intrinsic fractions, but show that a non-negligible sub-population could be present.}

\AH{Another caveat we take into account before interpreting our results is the probability to accidentally recover a false periodicity in our sample. Beyond the per-candidate, per-band bootstrap FAP reported in each candidate's subsection (Sect.~\ref{sec:global_fap}), App.~\ref{app:global_fap_estimate} gives an order-of-magnitude, survey-wide estimate of the expected number of false detections, and of the expected number of accidental dual-band detections specifically; Those effects are small compared to our three dual-band detections.}

We note that two of the three recovered periodicities are comparable to the 29.5-day lunar synodic period. The lunar cycle may, in principle, introduce spurious periodicities through the survey window function or varying sky background affecting image subtraction. Although we find no direct evidence linking our detections to this effect, we note it as a potential systematic that requires further investigation.

\AH{We assess the robustness of the three dual-band detections to the choice of polynomial degree in App.~\ref{sec:appendix_band_table}, alongside Table~\ref{tab:all_baselines}.}

The injection-recovery simulations demonstrate that the three detections should not be interpreted as a simple estimate, \(3/34\), of the intrinsic periodic fraction of SESNe. The probability of detecting a periodic modulation depends strongly on its period and amplitude, as well as on the cadence, photometric precision, and temporal baseline of each LC. In particular, our sensitivity is limited by the requirement that the observations span several cycles of the putative signal. Because we restrict the searched periods to \(P\lesssim T/2.5\), progressively longer signals can be tested in progressively fewer objects.

The present search is therefore primarily sensitive to relatively short periods, of order a few tens of days. Detecting periodicities of $\sim100$ days is observationally challenging, as it requires LCs spanning more than $\sim250$ days, by which time many SNe have either become seasonally unobservable or too faint for the survey detection threshold. The recovery map shown in Figure~\ref{fig:Pij_recovery_map} explores the period-amplitude range of the injection-recovery simulations, but periodic SESNe are not expected to be confined to this parameter space; see predictions in \citet{Ercolino+26} of periods extending up to $\sim10^3$ days. This survey is insensitive to such long periods.

The survey sensitivity alone is therefore insufficient to infer the intrinsic fraction of periodic SESNe, since the observed number of detections depends both on the period distribution of such events and their fraction in the population. By adopting a theoretical model for the intrinsic period distribution, however, the observed detections can be used to constrain the underlying fraction of periodic SESNe. We pursue this model-dependent comparison in the following section.

\subsection{Testing theoretical population models}
\label{subsec:ercolino_models}

To connect the observed detections with predictions for the underlying population, we combined our recovery map (Fig. \ref{fig:Pij_recovery_map}) with the population synthesis models of \cite{Ercolino+26}, which predict periodic variability arising from repeated CCI in binary systems that remain bound after the SN explosion.
Each model predicts both a distribution of post-SN periods for SN~2022jli-like systems and an intrinsic fraction of SESNe expected to exhibit such variability. Our analysis tests these two predictions jointly by determining whether the population predicted by each model would plausibly produce at least three detections when observed through the selection effects of our survey.

For each assumed number $X$ of intrinsically periodic SNe among the 34 dual-band objects, we generated Monte Carlo population realizations. Each simulated periodic event was assigned a period drawn from the period distribution of the model being tested and a semi-amplitude drawn uniformly between 0 and 0.2 mag. The event was then associated with the corresponding recovery probability, $P_{ij}$, measured in the injection-recovery simulations. A Bernoulli trial with success probability $P_{ij}$ determined whether that event was recovered. Summing these outcomes produced an integer number of detections, $N_{\rm det}$, for each realization.

Repeating this procedure allowed us to calculate $P(N_{\rm det}\geq3\mid X)$, where three is the observed number of dual-band detections. For each theoretical period distribution, we identify the smallest value of $X$ for which more than $5\%$ of realizations recover at least three detections. We denote the corresponding fraction by $f_5=X/34$. Intrinsic periodic fractions below $f_5$ reproduce the observed number of detections in fewer than $5\%$ of realizations and are therefore disfavored by our observations under the adopted model assumptions. Thus, $f_5$ is not an independent measurement of the intrinsic periodic fraction, but rather the minimum fraction compatible with observations, given the adopted theoretical period distribution and the sensitivity of our survey, as well as the assumed amplitude distribution. The population synthesis models of \cite{Ercolino+26} do not predict the observable photometric amplitudes, which may depend on additional factors such as the optical depth and geometry of the ejecta. Consequently, the models may over-predict the number of observable periodic events.

\AH{The adopted semi-amplitude ceiling of 0.2 mag (Sect.~\ref{sec:injection_recovery}) is chosen to comfortably span the semi-amplitudes of our three candidates ($0.04$--$0.20$ mag; Table~\ref{tab:double_hits_poly3}), rather than a physical prediction. To test the impact of this choice, we repeated the injection-recovery simulation for the $P=25$--$35$ day bin (which brackets the periods of SN~2020sgf and SN~2022esa) with the semi-amplitude extended up to 0.5 mag. Recovery does not saturate by 0.2 mag: it rises from $0.4$ at $A=0.2$ mag to $\approx0.5$, flattening only between $A=0.3$ and $0.5$ mag. Because our Monte Carlo population comparison draws amplitudes uniformly only up to 0.2 mag, real periodic events with larger amplitudes -- which are not excluded by any existing observational or theoretical constraint -- would be recovered more efficiently. Consequently, $f_5$ values reported below should be regarded as conservative (i.e., an upper limit on the minimum required intrinsic fraction): if the true amplitude distribution extends appreciably above 0.2 mag, a smaller intrinsic periodic fraction than $f_5$ could still plausibly reproduce our three detections.}

\AH{In Sect. \ref{subsec:2022jli} we show that our pipeline underestimates amplitudes of sawtooth-like periodic signals, like the one reported for SN~2022jli. Since the recovery efficienty grows with amplitude, this ensures our $f_5$ values remain conservative even after considering some targets might have a similar profile to SN~2022jli, and not a pure sinusoid.}

\AH{\citet{Ercolino+26} predict the occurrence and properties of CCI-driven periodicity using their population synthesis code SN-ORACLE \citep{Ercolino+26a}, built on a large grid of MESA single- and binary-star evolution models at galactic metallicity.} \citet{Ercolino+26} present 45 population synthesis models spanning three natal kick prescriptions, five merger criteria, and three explodability criteria. The model names encode these choices; for example, \texttt{k3m5e1} denotes the third kick prescription, fifth merger criterion, and first explosion criterion. \AH{The kick prescriptions (\texttt{k1}--\texttt{k3}) correspond to different natal-kick distributions, which differ mainly in the typical kick magnitude and thus in how often a binary becomes unbound after a SN versus bound and eccentric. The merger criteria (\texttt{m1}--\texttt{m5}) correspond to different prescriptions for the onset of the first, pre-SN Roche-lobe overflow episode, which set how many systems merge before the primary explodes. The explosion criteria (\texttt{e1}--\texttt{e3}) adopt different core-collapse explodability conditions, which determine which pre-SN stars are assumed to successfully explode (as opposed to collapsing directly to a black hole without a SN).} We selected seven representative models spanning the range of predicted periodic fractions. The resulting comparison is summarized in Table~\ref{tab:ercolino_models}.

The inferred minimum fractions differ between models because the result depends not only on the total number of periodic systems, but also on where those systems lie in period space. A population containing a larger proportion of short-period events is easier for our search to detect and can therefore reproduce three detections with a smaller intrinsic fraction. Conversely, a model in which most periodic systems occupy the long-period regime requires a larger total population to produce the same number of observable events.

\begin{table}
\centering
\caption{Comparison between the intrinsic periodic fractions predicted by the population synthesis models of \citet{Ercolino+26} (their Fig.~7) and the corresponding values of $f_5$ inferred from our simulations. Here, $f_5$ is the smallest intrinsic periodic fraction for which at least $5\%$ of realizations recover three or more detections (see Section~\ref{subsec:ercolino_models}).}
\label{tab:model_comparison}
\begin{tabular}{lccc}
\hline
Population Model & \AH{Predicted intrinsic fraction} & $f_5$ \\
\hline
k1m1e3 & 4\% & \AH{24\%} \\
k3m1e1 & 6\% & 18\% \\
k2m1e1 & 8\% & 18\% \\
k3m5e1 & 9\% & 21\% \\
k1m1e1 & 16\% & 24\% \\
k1m3e3 & 21\% & 21\% \\
k1m5e1 & 27\% & 21\% \\
\hline
\end{tabular}
\label{tab:ercolino_models}
\end{table}

Among the seven representative models, only \texttt{k1m3e3} and \texttt{k1m5e1} are not disfavored by our estimations. Interestingly, the \texttt{k1m1e1} model, identified by \citet{Ercolino+26} as their reference model, is only marginally disfavored, predicting an intrinsic periodic fraction of $15.6\%$, compared with $f_5=23.5\%$ inferred from our analysis. In other cases, the periodic fraction predicted by the model lies below the minimum fraction obtained using that model's own period distribution. The observations therefore indicate tension with each of the tested combinations of period distribution and intrinsic fraction. Under the adopted assumptions, these models predict too few periodic systems to make the recovery of three events sufficiently probable.

\subsection{Implications for the Population}

The three periodic SESNe identified in this work demonstrate that such events are unlikely to represent an exceptionally rare phenomenon. Although only a handful of periodically modulated SESNe have been reported to date, our population analysis suggests that they may constitute a non-negligible fraction of the SESN population. Under the theoretical period distributions considered here, the observations are compatible with intrinsic periodic fractions of order $\sim20\%$ or higher, indicating that periodically modulated SESNe could represent a \AH{non-negligible} sub-population rather than isolated peculiar events.

Our comparison with the population synthesis models of \citet{Ercolino+26} demonstrates that the observed detections are compatible with several currently available models. The agreement, however, depends not only on the predicted intrinsic periodic fraction, but also on the assumed intrinsic period and semi-amplitude distributions adopted in the simulations.

Future wide-field transient surveys with larger samples, denser cadence \citep[e.g. LAST;][]{Ofek+23, Ben-Ami+23}, and longer observational baselines will provide substantially stronger constraints on the population of periodically modulated SESNe and enable more robust tests of theoretical models.

\section{Summary} \label{sec:summary}

We have developed and applied a statistically rigorous pipeline to search for periodic modulation in SESN LCs. By fitting polynomial and sinusoidal components simultaneously and assessing detection significance through the improvement in $\Delta\chi^2$, the method provides a transparent and reproducible framework suitable for survey-scale searches. Applied to our sample, the pipeline successfully recovered the two previously reported periodic SNe, SN~2022jli and SN~2022esa, and identified SN~2020sgf as an additional promising periodic candidate. Another object SN 2019tsf is described in App. \ref{app:2019tsf}.

Using injection-recovery simulations, we quantified the sensitivity of our search and compared the observed detections with the population synthesis models of \citet{Ercolino+26}. Under the assumptions adopted in this work, models predicting intrinsic periodic fractions of order $\sim20\%$ are consistent with our observations\AH{, though we caution that this estimate rests on only three dual-band detections and should be regarded as illustrative rather than a precise measurement}. More generally, our results suggest that 2022jli-like events may constitute a non-negligible fraction of the stripped-envelope SN population rather than representing exceptionally rare objects.

\AH{The code used in this work is publicly available \footnote{\url{https://github.com/hoasaf3/sne_periodicity_search}} \citep{sesne_periodicity_search}, including all polynomial fits.}

\section*{Acknowledgments}
We thank Andreas Ercolino for generously sharing the population synthesis model data used in this work. We also thank Avshalom Badash and Erez A. Zimmerman for helpful discussions. \AH{We thank the anonymous referees for their valuable suggestions.}

E.O.O. is grateful for the support of grants from the Willner Family Leadership Institute, André Deloro Institute, Benoziyo center for astrophysics, the Orion center for ground-based astronomy, Paul and Tina Gardner,
The Norman E Alexander Family M Foundation ULTRASAT Data Center Fund, Israel Science Foundation, Israeli Ministry of Science, Minerva,
BSF, NSF-BSF, Israel Council for Higher Education (VATAT), Sagol Weizmann-MIT, Yeda-Sela, and the Rosa and Emilio Segr\`e Research Award.

AGY’s research is supported by ISF, IMOS and BSF grants, as well as the André Deloro Institute for Space and Optics Research, the Center for Experimental Physics, a WIS-MIT Sagol grant, the Norman E Alexander Family M Foundation ULTRASAT Data Center Fund, and Magnus Konow in honor of his mother Olga Konow Rappaport;  AGY is the incumbent of the The Arlyn Imberman Professorial Chair.

\AH{\software{Astropy \citep{astropy:2013, astropy:2018, astropy:2022}, astroplan \citep{astroplan}, NumPy \citep{numpy}, SciPy \citep{scipy}, Matplotlib \citep{matplotlib}, pandas \citep{pandas}}}

\appendix

\section{Baseline and Band-Level Diagnostics}
\label{sec:appendix_band_table}

In this Appendix we provide the full band-level periodicity diagnostics for the candidates discussed in Section~\ref{sec:results}. While the main text focuses on dual-band detections, at Table \ref{tab:all_baselines}  we report the results for each band, and for baselines of other polynomial degrees and not just 3.

For each supernova, we list the recovered best-fit period, amplitude, MDA, single-frequency $p$-value, and corresponding $\Delta\chi^2$ for all polynomial degrees considered (poly3–poly6) and for both photometric bands. These values correspond to the maximum $\Delta\chi^2$ obtained within the searched frequency range for each configuration.

This table is intended to provide full transparency regarding method dependence and band-level behavior. It presents the same three candidates listed in the dual-band table and SN~2019tsf presented in App. \ref{app:2019tsf}. It reports the detailed results for each band and for all polynomial baselines considered, rather than only the configurations that satisfied the dual-band consistency criterion.

\begin{table*}
\centering
\begin{tabular}{lccccccc}
\hline
SN & Band & Method & $P$ & $A$ & MDA & $p$-value & $\Delta\chi^2$ \\
& & & [Days] & [mag] & [mag] & & \\
\hline
\multirow{8}{*}{2022jli} & g & poly3 & 12.6 & 0.07 & 0.01 & $\AH{3}\times 10^{-12}$ & \AH{53} \\
& r & poly3 & 12.4 & 0.04 & 0.01 & \AH{$5\times 10^{-6}$} & \AH{25} \\
& g & poly4 & 12.6 & 0.07 & 0.01 & $2\times 10^{-13}$ & 59 \\
& r & poly4 & 12.4 & 0.04 & 0.01 & $\AH{3}\times 10^{-7}$ & \AH{30} \\
& g & poly5 & 12.6 & 0.07 & 0.01 & $2\times 10^{-13}$ & \AH{58} \\
& r & poly5 & 12.4 & 0.04 & 0.01 & $\AH{4}\times 10^{-7}$ & \AH{29} \\
& g & poly6 & \AH{12.4} & 0.07 & 0.01 & $3\times 10^{-13}$ & 58 \\
& r & poly6 & 12.4 & 0.04 & 0.00 & $\AH{3}\times 10^{-7}$ & \AH{30} \\
\hline
\multirow{8}{*}{2022esa} & g & poly3 & 34.2 & 0.14 & 0.03 & $\AH{1}\times 10^{-4}$ & 18 \\
& r & poly3 & 34.2 & 0.11 & \AH{0.01} & $5\times 10^{-5}$ & 20 \\
& g & poly4 & 36.3 & 0.15 & 0.02 & $5\times 10^{-5}$ & 20 \\
& r & poly4 & 36.3 & 0.12 & 0.01 & $2\times 10^{-5}$ & 21 \\
& g & poly5 & \AH{36.3} & 0.15 & 0.02 & $4\times 10^{-5}$ & 20 \\
& r & poly5 & 36.3 & 0.12 & 0.01 & $3\times 10^{-5}$ & 21 \\
& g & poly6 & 34.2 & 0.16 & 0.02 & $3\times 10^{-5}$ & 21 \\
& r & poly6 & \AH{36.3} & 0.12 & 0.01 & $\AH{5}\times 10^{-5}$ & 20 \\
\hline
\multirow{8}{*}{2020sgf} & g & poly3 & 30.2 & 0.20 & 0.02 & $6\times 10^{-10}$ & 42 \\
& r & poly3 & 35.9 & \AH{0.11} & 0.01 & $\AH{3}\times 10^{-9}$ & \AH{39} \\
& g & poly4 & 30.2 & 0.19 & 0.02 & $6\times 10^{-10}$ & 42 \\
& r & poly4 & 64.7 & 0.11 & 0.01 & $3\times 10^{-9}$ & \AH{39} \\
& g & poly5 & 28.9 & 0.15 & 0.02 & $\AH{4}\times 10^{-6}$ & 25 \\
& r & poly5 & 29.4 & 0.08 & 0.01 & $2\times 10^{-9}$ & 40 \\
& g & poly6 & 28.9 & 0.14 & 0.01 & $2\times 10^{-5}$ & \AH{21} \\
& r & poly6 & 29.4 & 0.07 & 0.01 & \AH{$1\times 10^{-8}$} & \AH{36} \\
\hline
\multirow{8}{*}{2019tsf} & g & poly3 & 55.4 & 0.24 & 0.03 & $1\times 10^{-4}$ & 18 \\
& r & poly3 & 56.6 & 0.16 & 0.03 & $2\times 10^{-3}$ & 12 \\
& g & poly4 & 55.4 & 0.21 & 0.03 & $3\times 10^{-4}$ & 16 \\
& r & poly4 & 50.4 & 0.11 & 0.02 & $\AH{5}\times 10^{-3}$ & 11 \\
& g & poly5 & 55.4 & \AH{0.20} & 0.02 & $4\times 10^{-4}$ & 16 \\
& r & poly5 & 47.9 & 0.09 & 0.02 & $\AH{9}\times 10^{-3}$ & 10 \\
& g & poly6 & 55.4 & 0.20 & 0.02 & $1\times 10^{-3}$ & 13 \\
& r & poly6 & 45.5 & 0.08 & 0.02 & $\AH{3}\times 10^{-2}$ & 7 \\
\hline
\end{tabular}
\caption{Single band periodicity diagnostics for dual-band detections, and SN~2019tsf mentioned in App. \ref{app:2019tsf}. For each polynomial baseline (poly3-poly6) and photometric band, we report the best-fit period, amplitude, minimal detectable amplitude, single-frequency $p$-value, and $\Delta\chi^2$. The periods per band in this Table were extracted using the global peak in $\Delta \chi^2$, unlike the harmonics-avoiding technique used for the rest of the pipeline (Sect. \ref{sec:model_fitting}) and provided here for the purpose of diagnostics. The typical errors in P are $\sim 5\%$.}
\label{tab:all_baselines}
\end{table*}

\AH{The three dual-band detections are not equally robust to the choice of polynomial degree, as the table below shows: SN~2022jli is recovered as a dual-band detection for every degree (poly3-poly6). SN~2020sgf is recovered for poly3, poly5, and poly6, but not poly4: its poly4 $g$-band period (30.2 d) is essentially unchanged from poly3, but its $r$-band best-fit period jumps to 64.7 d, a harmonic of the fundamental, so the two bands no longer agree within the period tolerance of Sect.~\ref{subsec:methods_significance}. SN~2022esa is recovered for poly4 and poly5, but not poly3 or poly6: at poly3 its $g$-band single-frequency $p$-value is $1.0\times10^{-4}$, at (rather than below) our threshold. At poly6 both bands are independently significant with consistent periods, but the narrow period-tolerance window around the anchor period happens to exclude the interior local maximum required by our matching procedure. In every disfavored case the underlying $\Delta\chi^2$ remains high and the change in classification reflects a specific, identifiable boundary effect (a harmonic alias, a threshold value, or a tolerance-window edge) rather than a qualitative loss of significance. SN~2020sgf's $r$-band residuals are also the one case among the six band-level fits with a marginal deviation from Gaussianity (App.~\ref{app:shapiro_wilk}), consistent with its reduced robustness here. Finally, App.~\ref{sec:gls_comparison} shows that a sequential detrend+Lomb--Scargle analysis recovers a different global-best period for SN~2020sgf. Taken together, these checks indicate that SN~2022jli is a robust detection under every variation we tested, while SN~2020sgf and SN~2022esa each sit close enough to our significance threshold that they are sensitive, in different specific ways, to defensible variations in the analysis choices. We retain all three dual-band detections in our fiducial ($d=3$) results, but the population inference in Sect.~\ref{subsec:ercolino_models} should be read with this reduced robustness for two of the three detections in mind.}

\subsection{\AH{Comparison with a sequential detrend + Lomb--Scargle analysis}}
\label{sec:gls_comparison}

\AH{We compared our joint fit with the common sequential approach for the three dual-band detections: we subtracted the same 3rd-order polynomial baseline and ran a Lomb--Scargle periodogram \citep[LSP; via \texttt{astropy};][]{astropy:2022} on the residuals over the same frequency grid, with the bootstrap FAP of Sect.~\ref{sec:global_fap} (Table~\ref{tab:gls_comparison}).}

\AH{The two methods agree for SN~2022jli and SN~2022esa. For SN~2020sgf, the LSP's best period ($\approx56$--$57$ d) is a harmonic of the joint method's $\approx30$--$36$ d detection; in both methods the two peaks are nearly tied (the weaker reaching $\geq89\%$ of the stronger), further indicating that this detection is less robust than SN~2022jli's.}

\begin{table}
\centering
\caption{\AH{Comparison between our joint poly+sinusoid method and a sequential detrend+Lomb--Scargle (LSP) analysis, for the three dual-band detections. $P_{\rm joint}$ and $P_{\rm LSP}$ are the recovered periods; FAP$_{\rm LSP}$ is the LSP's bootstrap false-alarm probability, computed with the same methodology as Sect.~\ref{sec:global_fap}.}}
\label{tab:gls_comparison}
\begin{tabular}{lcccc}
\hline
SN & Band & $P_{\rm joint}$ [d] & $P_{\rm LSP}$ [d] & FAP$_{\rm LSP}$ \\
\hline
\multirow{2}{*}{2022jli} & $g$ & 12.56 & 12.56 & $<5\times10^{-4}$ \\
 & $r$ & 12.41 & 12.41 & $<5\times10^{-4}$ \\
\hline
\multirow{2}{*}{2022esa} & $g$ & 34.24 & 32.39 & $<5\times10^{-4}$ \\
 & $r$ & 34.21 & 34.21 & $<5\times10^{-4}$ \\
\hline
\multirow{2}{*}{2020sgf} & $g$ & 30.17 & 56.72 & $<5\times10^{-4}$ \\
 & $r$ & 35.94 & 55.77 & $<5\times10^{-4}$ \\
\hline
\end{tabular}
\end{table}

\AH{Nonetheless, the sequential approach can bias the amplitude low. In a toy model of a quadratic baseline plus a sinusoid spanning $T/P=2.5$ cycles (our minimum; Sect.~\ref{sec:model_fitting}), both approaches recover the injected period, but the sequential one underestimates the amplitude by $\approx11\%$, whereas the joint fit is unbiased. This is expected from the Frisch--Waugh--Lovell theorem \citep{Frisch+33, Lovell63}: the joint fit is equivalent to detrending both the data and the sinusoidal regressors, while the sequential approach detrends only the data, so the part of the sinusoid that the polynomial can mimic is absorbed by the baseline.}

\subsection{\AH{Gaussianity of the fit residuals}}
\label{app:shapiro_wilk}

\AH{As a check on the Gaussianity assumption underlying the $\chi^2$ formalism (Sect.~\ref{subsec:methods_significance}), we ran a Shapiro--Wilk test on the standardized baseline-fit residuals of all six band-level fits in Table~\ref{tab:band_level}: five are consistent with Gaussianity ($p>0.1$), while SN~2020sgf's $r$-band residuals show a marginal deviation ($p=0.009$).}

\subsection{\AH{Order-of-magnitude survey-wide false-alarm estimate}}
\label{app:global_fap_estimate}

\AH{Sect.~\ref{sec:global_fap} assesses the significance of each candidate empirically, band by band, via a bootstrap test. As a complementary, order-of-magnitude check on the survey as a whole, we also estimate the expected number of false detections analytically. The $p$-value defined in Eq.~\ref{eq:pval} corresponds to a single trial frequency. However, for each light curve we scan a range of frequencies, and the survey includes multiple light curves, so the probability of obtaining at least one false detection increases with the number of independent trials.}

\AH{For a given light curve with time baseline $T$, the approximate number of independent frequencies is}
\begin{equation}
M \sim 2 (f_{\max} - f_{\min})\, T,
\end{equation}
\AH{which reflects the Nyquist frequency $1/2T$ across the searched frequency interval. If $p_{\mathrm{single}}$ denotes the single-frequency $p$-value threshold, the false-alarm probability for that light curve is approximately}
\begin{equation}
p_{\mathrm{LC}} = 1 - (1 - p_{\mathrm{single}})^M
\approx M\,
p_{\mathrm{single}},
\label{eq:p_lc}
\end{equation}
\AH{valid when $p_{\mathrm{single}} \ll 1$. For a survey of $N_{\mathrm{LC}}$ independent light curves, the expected total number of false detections is then}
\begin{equation}
N_{\mathrm{false}} \approx
\sum_{i=1}^{N_{\mathrm{LC}}}
M_i\, p_{\mathrm{single}},
\label{eq:global_fap}
\end{equation}
\AH{where $M_i$ accounts for the individual time baseline of each light curve; the time baselines in our sample vary from $\sim$10 to 250 days. With $p_\mathrm{single}=10^{-4}$, we find $N_\mathrm{false}\approx0.8$ across all 117 analyzed light curves, and $N_\mathrm{false}\approx0.5$ for the 68 light curves belonging specifically to the 34 dual-band candidates. Assuming, for illustration, that false alarms in the two bands of a given candidate are independent, this corresponds to an expected $34\times(0.5/68)^2\sim2\times10^{-3}$ accidental dual-band detections, before imposing the additional requirement that the recovered periods agree; although the two bands are not strictly independent, this illustrates the strong suppression of false positives achieved by the dual-band criterion.}

\AH{This estimate assumes independent trials on a regular, dense frequency grid, which irregular sampling can violate. It is therefore only an order-of-magnitude complement to the per-band bootstrap test.}

\section{Another possible candidate}
\label{app:2019tsf}

\begin{figure}
    \centering
    \includegraphics[width=\linewidth]{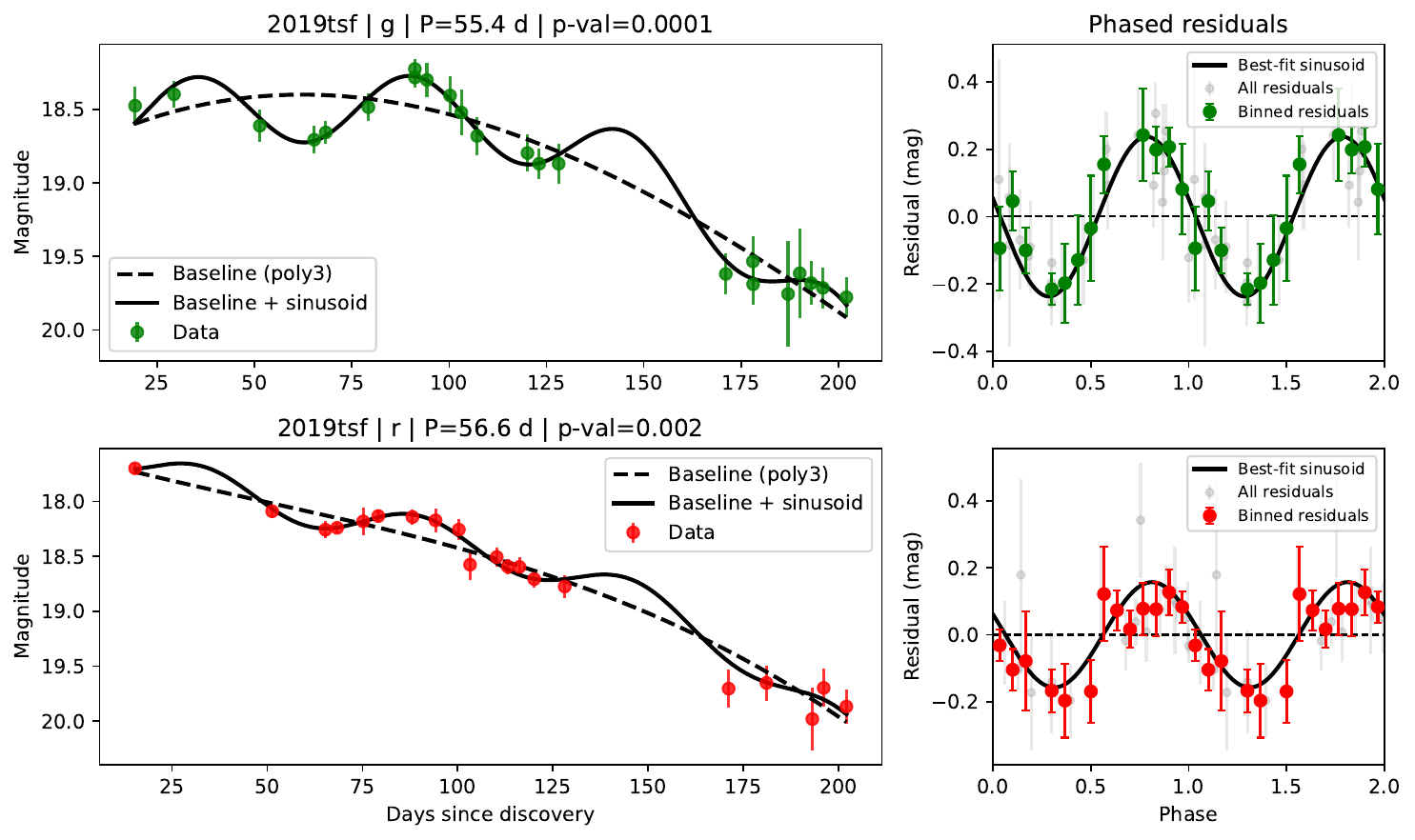}
    \caption{Like Fig. \ref{fig:2020sgf_fit} but for SN 2019tsf.}
    \label{fig:2019tsf_fit}
\end{figure}

\begin{figure}
    \centering
    \includegraphics[width=\linewidth]{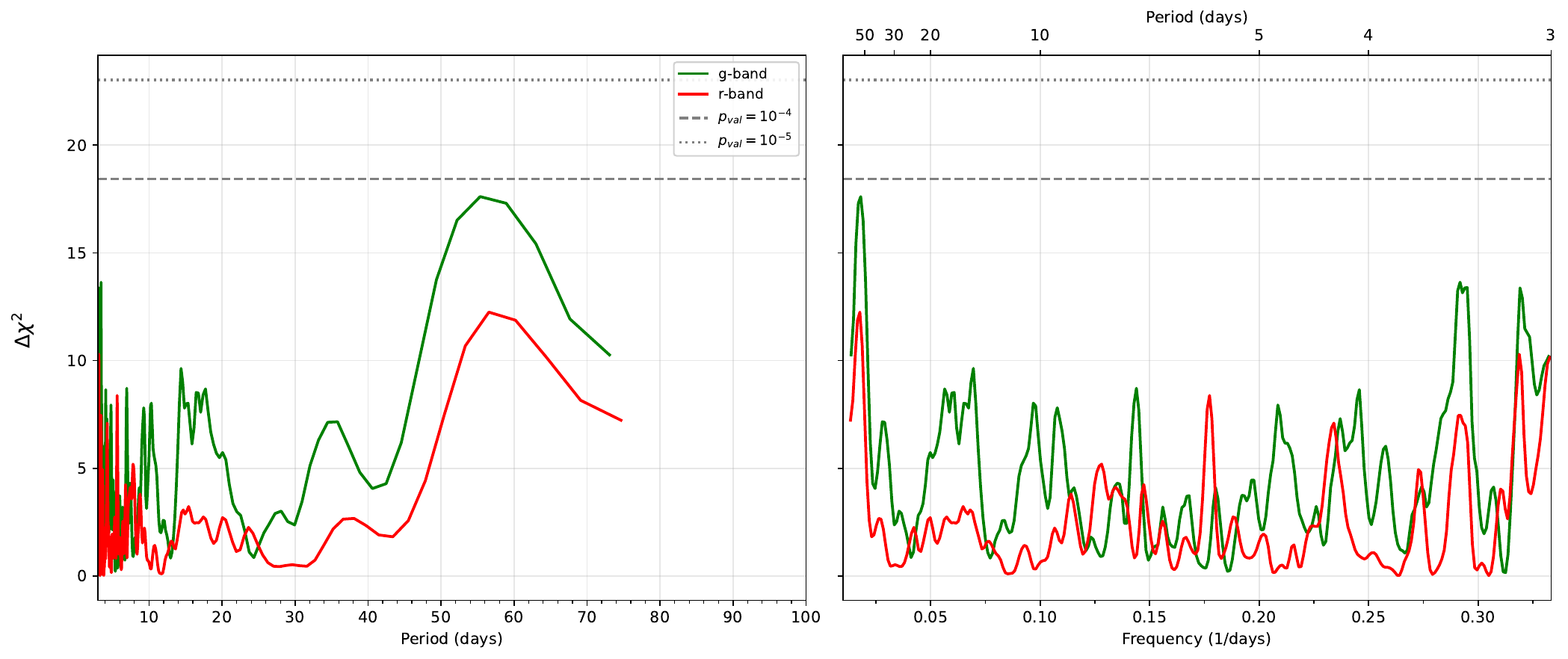}
    \caption{Same as Fig. \ref{fig:2022jli_periodogram} but for SN 2019tsf. Peaks are recovered around $P \approx 55$ days for both $g-$ and $r-$ band, but not exceeding our significance threshold (lower dashed grey line).}
    \label{fig:2019tsf_periodogram}
\end{figure}

SN 2019tsf is a Type Ib SN that caught our attention while making this search. This object has been studied \citep[e.g.][]{Sollerman+20, Zenati+25} for having multiple peaks in its light curve. It did not pass our strict filters for a detection and is therefore not counted for in our population statistics. We do, however, believe it is worth mentioning. Its light curve is shown in Fig. \ref{fig:2019tsf_fit}, where the inclusion of a sinusoidal component yields a preferred period of $\approx 55$ days in both bands. The improvement over the polynomial-only baseline is non-negligible in both photometric bands. We note, however, that the light curve contains an observational gap of approximately 50 days, comparable to the inferred period, which may affect the recovered period. The periodogram of SN~2019tsf is shown in Fig. \ref{fig:2019tsf_periodogram}, where both peaks at $\approx 55$ days are visible, however neither exceed our criterion for significance. \AH{A bootstrap FAP test (as described in Sect.~\ref{sec:global_fap}) gives FAP~$<5\times10^{-4}$ in the $g$ band and FAP~$=2\times10^{-3}$ in the $r$ band.}

\bibliography{refs}{}
\bibliographystyle{aasjournalv7}

\end{document}